\documentclass[preprint,3p]{elsarticle}%
\usepackage{amssymb}
\usepackage{amsthm}
\usepackage{mathrsfs} 
\usepackage[mathlines]{lineno}
\usepackage{graphicx}
\usepackage{units}
\usepackage{url}
\usepackage{ulem}
\biboptions{sort&compress}
\usepackage{amsmath}
\usepackage{amsfonts}
\usepackage{bm}
\usepackage{textcomp}
\usepackage{subfigure}
\usepackage{multicol}
\usepackage{verbatim}
\usepackage{rotating}
\usepackage[colorlinks,linkcolor=red,citecolor=red]{hyperref}
\usepackage{float}
\usepackage{epsfig}
\usepackage{dcolumn}
\usepackage{bm}
\usepackage{color}
\usepackage{pstricks}
\usepackage{pst-node}
\usepackage[T1]{fontenc}
\usepackage{times}
\usepackage{indentfirst}
\usepackage[english]{babel}
\usepackage{epstopdf}
\addto{\captionsenglish}{%

}

\journal{Physics Letters B}
\makeatletter
\newenvironment{tablehere}
{\def\@captype{table}}
{}

\newenvironment{figurehere}
{\def\@captype{figure}}
 {}
\newcommand{\bcl}{\begin{center}}
\newcommand{\ecl}{\end{center}}

\newcommand{\EE}{e^+e^-}

\def \ee   {e^+e^-}

\def \mev  {\mbox{MeV}}

\begin{document} 
\begin{frontmatter} 
\title{\bf \boldmath Search for invisible decays of a dark photon using $\EE$ annihilation data at BESIII} 
\author{
\small
\texorpdfstring{
M.~Ablikim$^{1}$, M.~N.~Achasov$^{11,b}$, P.~Adlarson$^{70}$, M.~Albrecht$^{4}$, R.~Aliberti$^{31}$, A.~Amoroso$^{69A,69C}$, M.~R.~An$^{35}$, Q.~An$^{66,53}$, X.~H.~Bai$^{61}$, Y.~Bai$^{52}$
% ...
\\
\vspace{0.2cm}
(BESIII Collaboration)\\
\vspace{0.2cm} {\it
$^{1}$ Institute of High Energy Physics, Beijing 100049, People's Republic of China\\
% ...
$^{4}$ Bochum Ruhr-University, D-44780 Bochum, Germany\\
% ...
$^{11}$ G.I. Budker Institute of Nuclear Physics SB RAS (BINP), Novosibirsk 630090, Russia\\
% ...
$^{31}$ Johannes Gutenberg University of Mainz, Johann-Joachim-Becher-Weg 45, D-55099 Mainz, Germany\\
% ...
$^{35}$ Liaoning Normal University, Dalian 116029, People's Republic of China\\
% ...
$^{52}$ Southeast University, Nanjing 211100, People's Republic of China\\
$^{53}$ State Key Laboratory of Particle Detection and Electronics, Beijing 100049, Hefei 230026, People's Republic of China\\
% ...
$^{61}$ University of Jinan, Jinan 250022, People's Republic of China\\
% ...
$^{66}$ University of Science and Technology of China, Hefei 230026, People's Republic of China\\
% ...
$^{69}$ University of Turin and INFN, (A)University of Turin, I-10125, Turin, Italy; (B)University of Eastern Piedmont, I-15121, Alessandria, Italy; (C)INFN, I-10125, Turin, Italy\\
$^{70}$ Uppsala University, Box 516, SE-75120 Uppsala, Sweden\\
% ...
$^{b}$ Also at the Novosibirsk State University, Novosibirsk, 630090, Russia\\
% ...
}
\vspace{0.4cm}}{}
}

\date{\today} 
\begin{abstract}
  We report a search for a dark photon using $14.9$~fb$^{-1}$ of $\ee$
  annihilation data
  taken at center-of-mass energies from 4.13 to 4.60~GeV with the
  BESIII detector operated at the BEPCII storage ring. The dark photon
  is assumed to be produced in the radiative annihilation process of $\EE$ and
  to predominantly decay into light dark matter particles, which
  escape from the detector undetected.  The mass range from 1.5 to
  2.9~GeV is scanned for the dark photon candidate, and no
  significant signal is observed. The mass dependent upper limits at
  the 90$\%$ confidence level on the coupling strength parameter
  $\epsilon$ for a dark photon coupling with an ordinary photon vary
  between $1.6\times 10^{-3}$ and $5.7\times10^{-3}$.
  \end{abstract} 

\begin{keyword} 
 dark sector \sep dark photon \sep invisible decays 
\end{keyword} 
\end{frontmatter}

\begin{multicols}{2}
\section{Introduction}
The Standard Model (SM) of particle physics is powerful but does not
address several important phenomena that hint there is physics beyond
the SM.  One  that is not included in the SM is the existence of dark
matter (DM), which makes up $\sim$84$\%$ of the matter in the
universe~\cite{Planck:2018vyg,Bertone:2016nfn}.  DM
interactions with ordinary matter are observed through the
gravitational effects DM has on galaxies, but the lack of interaction
between DM and SM particles via strong, weak, and
electromagnetic forces makes it very challenging to detect directly in
particle physics experiments.
However, recent results from astrophysical observations~\cite{PAMELA:2008gwm,
  Fermi-LAT:2011baq, AMS:2013fma}, as well as the long-standing
discrepancy between the experimental value and the theoretical
prediction of the muon anomalous magnetic
moment~\cite{Muong-2:2006rrc,Muong-2:2021ojo} indicate there could be
a new force between the dark sector and the SM. The new force could be
mediated by a $U(1)_{D}$ gauge boson $\gamma'$ (referred to as a dark
photon), which couples weakly to a SM photon through kinetic mixing
($\frac{1}{2}\epsilon F_{\mu\nu}'F^{\mu\nu}$)
~\cite{Holdom:1985ag, Essig:2013lka,Arkani-Hamed:2008hhe,Essig:2009nc,Andreas:2012mt},
where $F_{\mu\nu}'$ and $F_{\mu\nu}$ are the field strengths of the
dark photon and the SM photon, respectively, and the mixing parameter
$\epsilon$ gives the coupling strength between the dark photon and SM
photon.

A dark photon with mass in the GeV range and $\epsilon$ value as high
as $\epsilon\sim$$10^{-3}$ has been predicted in
Refs.~\cite{Essig:2013lka,Arkani-Hamed:2008kxc,Pospelov:2007mp}.  The
small value of epsilon and a suppression factor of $\epsilon^2$ of the
coupling between the dark photon and the SM photon make it very hard
to be hunted.  However, benefiting from high intensity facilities such as
$\EE$ storage rings, a dark photon candidate could be produced in
particle physics experiments.  The dark photon would predominately
decay into a pair of DM particles ($\gamma'\to\chi\bar{\chi}$), which
are the lightest DM particles ($\chi$) with masses
$m_{\chi}<m_{\gamma'}/2$ and thus be invisible.  Previous measurements
on the invisible $\gamma'$ decays have been performed by the
NA62~\cite{NA62:2019meo}, NA64~\cite{NA64:2016oww,Andreev:2021fzd},
BaBar~\cite{BaBar:2017tiz}, E787~\cite{E787:2001urh} and
E949~\cite{BNL-E949:2009dza} experiments.  No evidence is found for
a dark photon, and upper limits for $\epsilon$ have been set.

A dark photon can also be searched for at the BESIII experiment in the 
radiative annihilation process $\EE\to\gamma \gamma'$, followed by an invisible decay of the $\gamma'$.
The dark photon candidate of mass $m_{\gamma'}$ would
be signified by the presence of a monochromatic photon with energy
\begin{linenomath}
\begin{equation}
\label{eq:eisr}
E_{\gamma}=\frac{s-m_{\gamma'}^2}{2\sqrt{s}},
\end{equation}
\end{linenomath}
where $\sqrt{s}$ is the $e^+e^-$ center-of-mass (c.m.) energy.
Compared with the BaBar measurement, BESIII runs at a relative lower c.m. energy
and can explore the low $m_{\gamma'}$ region with a finer binning scheme.
A novel method by exploiting the muon detector to veto the dominant QED background
is also developed and the purity of the final data sample is well controlled which is necessary
for this search.

In this letter, we perform a measurement of the process
$\EE\to\gamma \gamma'$ at the BESIII experiment located at
the Beijing Electron Positron Collider
(BEPCII)~\cite{Yu:IPAC2016-TUYA01}, where the dark photon decays
invisibly into a dark matter $\chi\bar{\chi}$ pair. Therefore, the
signal process only contains monochromatic single photon events.
We reconstruct the dark photon signal by seeking a narrow peak
in the photon energy ($E_{\gamma}$) spectrum, which is related to
the missing mass $m_{\gamma'}$ of the recoil system through
Eq.~(\ref{eq:eisr}).  The analysis is based on the data sets taken at
$\sqrt{s}=4.13$ to $4.60$~GeV, corresponding to an integrated
luminosity of 14.9~fb$^{-1}$~\cite{BESIII:2022xii}.

\section{Detector and Monte Carlo simulation}

The BESIII detector~\cite{Ablikim:2009aa} records symmetric $e^+e^-$
collisions provided by the BEPCII storage
ring~\cite{Yu:IPAC2016-TUYA01}, which operates with a peak luminosity
of $1\times10^{33}$~cm$^{-2}$s$^{-1}$ in the center-of-mass energy
range from 2.0 to 4.95~GeV.  BESIII has collected large data samples
in this energy region~\cite{Ablikim:2019hff}. The cylindrical core of
the BESIII detector covers 93\% of the full solid angle and consists
of a helium-based multilayer drift chamber~(MDC), a plastic
scintillator time-of-flight
 system~(TOF), and a CsI(Tl) electromagnetic calorimeter~(EMC), which
 are all enclosed in a superconducting solenoidal magnet providing a
 1.0~T magnetic field.
The solenoid is supported by an octagonal flux-return yoke with 9 layers of resistive
 plate counters interleaved with steel, comprising a muon
 identification system (MUC).
The
 charged-particle momentum resolution at $1~{\rm GeV}/c$ is $0.5\%$,
 and the specific energy loss (${\rm d}E/{\rm d}x$) resolution is $6\%$ for electrons from
 Bhabha scattering. The EMC measures photon energies with a resolution
 of $2.5\%$ ($5\%$) at $1$~GeV in the barrel (end cap) region. The
 time resolution in the TOF barrel region is 68~ps, while that in the
 end cap region is 110~ps. The end cap TOF system was
 upgraded in 2015 using multigap resistive plate chamber technology,
 providing a time resolution of
 60~ps~\cite{etof}.  The
spatial resolution in the MUC is better than 2~cm.

Simulated Monte Carlo (MC) samples produced with {\sc
  geant4}-based~\cite{geant4} software, including the geometrical
description of the BESIII detector and the detector response, are used
to determine the detection efficiency, and to estimate potential
backgrounds.  The signal MC events for the reaction
$e^+e^-\to\gamma \gamma'$ are generated using {\sc
  evtgen}~\cite{ref:evtgen} for 29 different $\gamma'$ mass hypotheses
in the range from 1.5 to 2.9~GeV with a 50~MeV step size. The possible background
sources are investigated with inclusive MC simulation samples,
consisting of open-charm processes, the Initial-State-Radiation production of lower mass
vector charmonium(-like) states, the continuum processes $e^+e^-\to q\bar{q} ~(q = u, d, s)$, 
and the QED processes $e^+e^-\to (\gamma) e^+e^-, (\gamma)\mu^+\mu^-, (\gamma)\gamma\gamma$. The
known decay modes of charmed hadrons are handled by 
{\sc evtgen}~\cite{ref:evtgen} with known decay branching fractions taken
from the Particle Data Group (PDG)~\cite{pdg}, and the remaining
unknown decays with {\sc lundcharm}~\cite{ref:lundcharm}.
The QED processes are generated with the {\sc babayaganlo}
generator~\cite{Balossini:2006wc}.

\section{Event selection}
The signal events have only one monochromatic photon. Thus, events
with any reconstructed charged tracks in the MDC are rejected.  Photon
candidates are required to have deposited energy larger than 25~$\mev$
in the barrel EMC region ($|\!\cos\theta|<0.8$) or larger than
50~$\mev$ in the end cap region ($0.86<|\!\cos\theta|<0.92$), where
$\theta$ is the polar angle of each photon candidate.  To remove
contamination from fake photons (due to electronic noise of the EMC), the number of photons ($N_\gamma$) in
each event should satisfy $1\le N_\gamma \le 3$. The
most energetic photon is regarded as the candidate for the signal
photon. 
To suppress the overwhelming di-gamma background events, the polar angle of the signal photon is required to
be within $|\!\cos\theta_{\gamma}|<0.6$.  
As our signal events are purely neutral events, the event start time $t_0$ is determined from hits in TOF when available or otherwise the trigger time by the EMC. 
To suppress electronic noise and energy deposition unrelated to the
physical events, the EMC time of the signal photon is required to be
within [-500, 1250]~ns with respect to $t_0$, where the values are
obtained by studying a di-gamma control sample.


Backgrounds with multiple photons in the final state, such as $\EE\to\gamma\gamma,~\gamma\gamma\gamma$, etc., can be effectively suppressed by requiring the total deposited energy of showers except for the signal photon (denoted as $E_{\rm extra}$) in the EMC be less
than 80~MeV.  To eliminate the backgrounds from neutral hadrons, such
as $e^{+}e^{-}\to n\bar{n}$ which also produce showers in the EMC,
two shower shape related variables, the lateral moment and energy
ratio in $3\times3$ and $5\times5$ crystals around the central seed crystal ($E9/E25$), are used to
distinguish showers caused by neutral hadrons from photons. The
lateral moment is given by 
\begin{linenomath}
\begin{equation} 
\frac{\Sigma_{i=3}^n E_ir_i^2}{\Sigma_{i=3}^n E_ir_i^2 +E_1r_0^2+E_2r_0^2},
\end{equation}
\end{linenomath}
where $n$ is the number of crystals associated with the shower, $E_i$ is the
deposited energy in the $ith$ crystal and $E_1 > E_2 > \ldots > E_n$,
$r_i$ is the lateral distance between the central and the $ith$
crystal, and $r_0$ is the average distance between two crystals.
The signal photon candidate is required to have a lateral moment
larger than 0.14 and less than 0.3 as well as a value of $E9/E25>0.95$,
which is optimized by studying the hadron background events from inclusive MC sample.

Due to the structure of the EMC, one of the photons from a di-gamma
event can penetrate the calorimeter in the gap between the barrel and
the endcap ($0.8<|\!\cos\theta|<0.86$), or occasionally pass between
crystals in the zenith angle direction ($\cos\theta=0$), producing a
sole photon in the detector, and thereby mimicking a signal event. An
escaped energetic photon often interacts in the  outer detector
material and produces secondary particles, which are then
recorded by the MUC.  To suppress this background, events with MUC
hits within $|\cos\theta_{\rm hit}|>0.65$ or $|\!\cos\theta_{\rm
  hit}|<0.06$, where $\theta_{\rm hit}$ is the polar angle of the MUC
hit, are rejected.

\section{The signal photon energy spectrum}
\begin{figurehere}
\bcl
        \centering
	\subfigure{\includegraphics[width=0.45\textwidth]{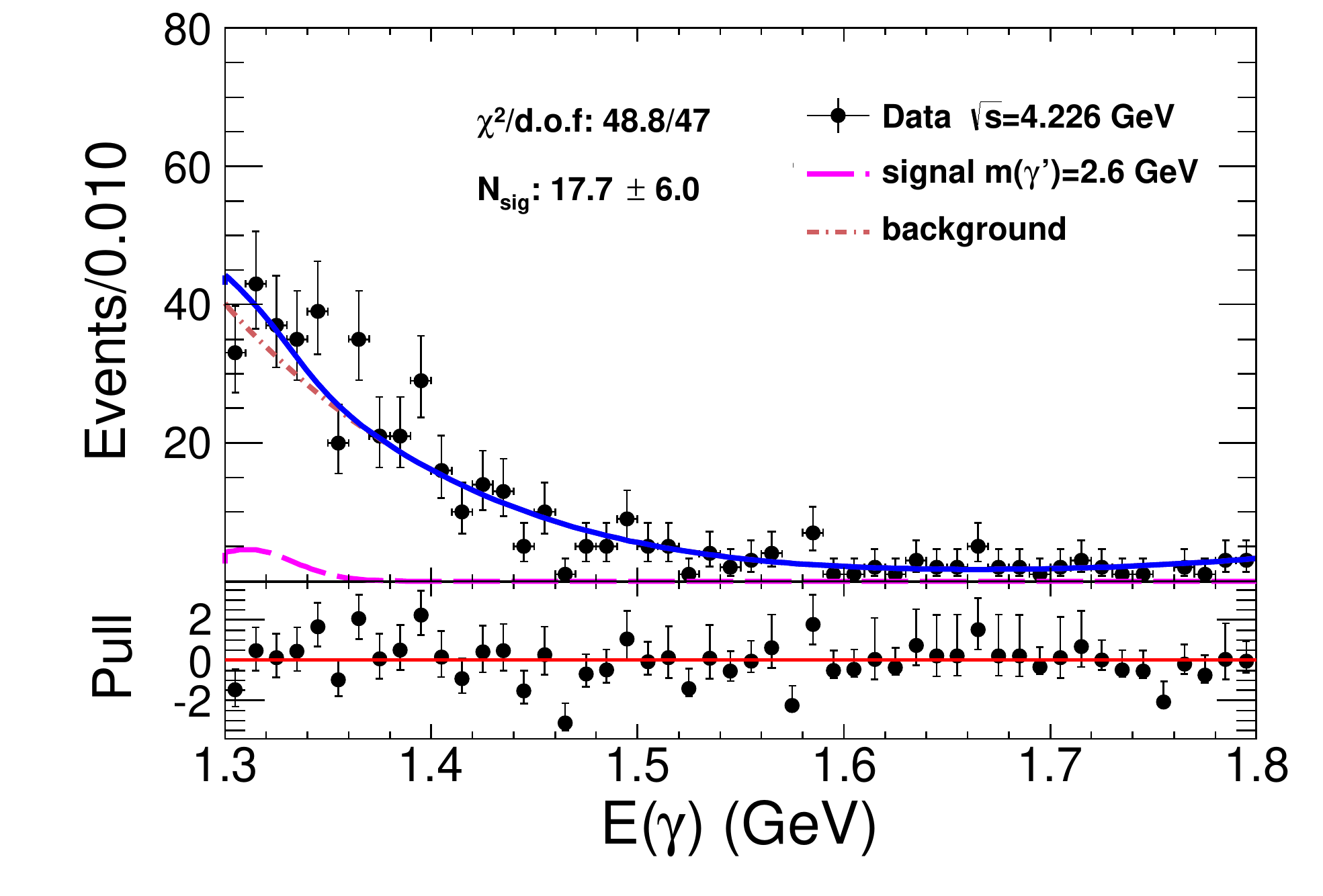}}
	\subfigure{\includegraphics[width=0.45\textwidth]{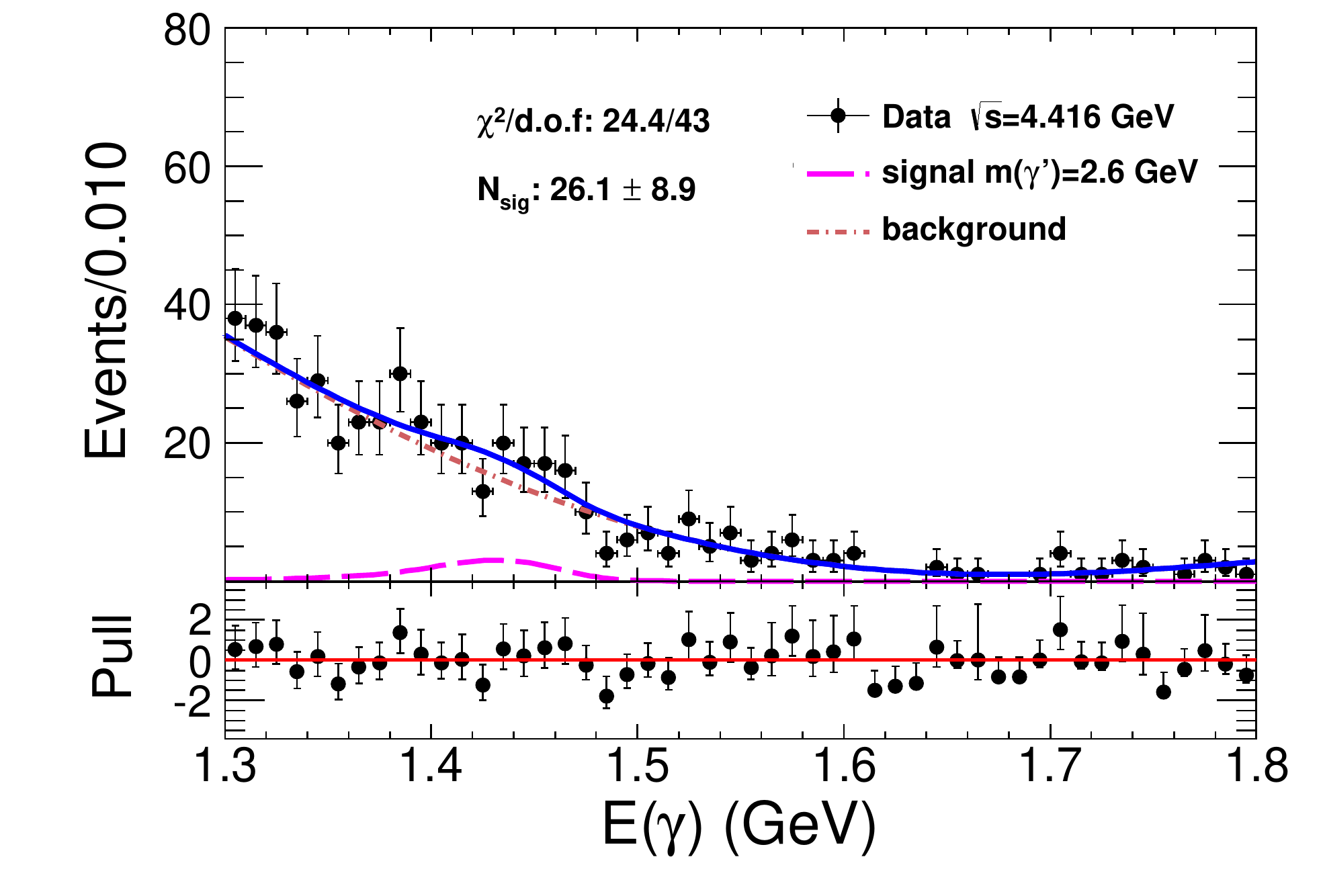}}
	\caption{The signal photon energy spectra at $\sqrt{s}$ = 4.226~GeV (top) and $\sqrt{s}=$4.416~GeV (bottom) with fit projections overlaid. The magenta dashed curves are the dark photon signal shapes with $m_{\gamma'}$ = 2.6~GeV, which depend on
	the c.m.\ energy of the data set according to Eq.~(\ref{eq:eisr}).
	Dots with error bars are data, the blue solid curves are the total fit results, and the red dot-dashed curves describe the background contributions.}
	\label{fig:fit_res}	
\ecl	
\end{figurehere}

After imposing the above event selection criteria, the signal photon energy spectra at $\sqrt{s}=4.178$ and 4.226~GeV are shown as
examples in Fig.~\ref{fig:fit_res}.
The trigger efficiency for single photon with $E(\gamma)<1.3$~GeV in the barrel EMC is
relatively low, and the background from di-gamma events is also high for the low energy region. 
In addition, a photon candidate with energy larger than 2~GeV will
saturate the EMC electronics, which results in shower loss.  Such
effects would lead to a significant number of di-gamma background
events when one energetic photon candidate is lost.  Therefore, we
only search for a dark photon signal within the range
$1.3<E(\gamma)<1.8$~GeV corresponding to
$1.5<m_{\gamma'}<2.9$~GeV, according to Eq.~(\ref{eq:eisr}).

 The production cross section for a
dark photon candidate can be calculated by~\cite{Essig:2009nc}
\begin{linenomath} \begin{equation}\label{xs} \centering \begin{split}
&\sigma(\EE\to\gamma \gamma')=\\ &
\frac{2\pi\alpha^2}{s}\epsilon^{2}(1-x)\left(
{\Big(1+\frac{2x}{(1-x)^2}\Big)\Theta-2\cos\theta_{\rm c}}\right), \\
  \end{split}
\end{equation}
\end{linenomath}
where $x=\frac{m^2_{\gamma'}}{s}$,
$\Theta=\log\frac{(1+\cos\theta_{\rm c})^2}{(1-\cos\theta_{\rm
    c})^2}$, $\cos\theta_{\rm c}=0.6$ is the $\cos\theta$ limit for
the signal photon polar angle, and $\alpha$ is the fine-structure
constant. The coupling strength parameter $\epsilon^2$ which determines
the cross section is common and shared for all c.m.~energy data
samples. The expected signal yield at a certain c.m.~energy is $N^{\rm
  sig}=\mathcal{L}\sigma(\EE\to\gamma \gamma')\mathscr{\epsilon_{\rm det}}\mathscr{\epsilon_{\rm
    trig}}$, where $\mathcal{L}$ is the integrated luminosity of the
data~\cite{BESIII:2022xii}, $\mathscr{\epsilon_{\rm det}}$ is the
signal detection efficiency and $\mathscr{\epsilon_{\rm trig}}$ is the trigger
efficiency.

The detection efficiencies for signal events at each c.m.\ energy for specific
values of $m_{\gamma'}$ are obtained by simulated signal MC samples, and
depending on $m_{\gamma'}$ and the c.m.\ energy, vary between 1\% and 6\%.
The trigger system at BESIII combines measurements from the MDC,
EMC, as well as TOF detectors to suppress beam-related background
and to record $\EE$ collision events as much as possible.  
The trigger condition valid for single photon events with $|\cos\theta_\gamma|<0.6$ only consists of EMC
information, which requires an event at least
has one reconstructed shower in the barrel region of the EMC, 
with a total deposit energy greater than 650 MeV~\cite{Berger:2010my}.
The corresponding trigger efficiency for this condition 
is studied via radiative Bhabha events ($\ee\to\gamma\ee$) 
using a same method as described in Ref.~\cite{Berger:2010my}. 
In order to select a
clean radiative Bhabha sample in the EMC barrel region, we require
either $e^\pm\gamma$ or $\EE$ be detected by the endcap detector and a
third $e^\mp$ or $\gamma$ hit the barrel EMC. We find
$\mathscr{\epsilon_{\rm trig}}=(99.40\pm0.01)\%$ for events with
$E(\gamma)>1.3$~GeV, and the efficiency drops dramatically
for $E(\gamma)<1.3$~GeV.

To determine the signal yield for each $m_{\gamma'}$ hypothesis, a
simultaneous unbinned maximum likelihood fit to the photon energy spectra is
performed to all data sets. 
The signal yields for each data set depend on the common parameter $\epsilon$, according to Eq.~(\ref{xs}). 
In the fit, the signal probability density function (PDF) for each c.m.\ energy data 
is modelled by a templated shape constructed from
the corresponding signal MC simulation
convolved with a Gaussian function, which represents the photon energy resolution
difference between data and MC simulation. The parameters of the
Gaussian function are determined by a di-gamma control sample.  The
background shape for each c.m.~energy data is described by a $4^{th}$-order 
Chebychev polynomial function with free parameters to fit data.
To obtain the signal yield for different
$m_{\gamma'}$ candidates, we scan the $m_{\gamma'}$ region with a 50
MeV step size (in total 29 steps) and repeat the
fit. Figure~\ref{fig:fit_res} shows the fit results for a dark photon
candidate with $m_{\gamma'} = 2.6$ GeV at $\sqrt{s} = 4.26$ and 4.416
GeV.  After taking into account the uncertainty from the background model, the
maximum local significance is determined to be $3.1 \sigma$ at
$m_{\gamma'} = 2.6$ GeV.  The statistical significances are calculated
by comparing the likelihoods with and without the signal components
in the fit, and taking the change of the number of degrees of freedom
into account.  Figure~\ref{fig:sig} shows the estimated local
statistical significance for different $m_{\gamma'}$ hypotheses.  The
maximum global significance, taking the Look-Elsewhere
Effect~\cite{BaBar:2012wey} into account, is determined to be
2.2$\sigma$.  Therefore, a null result is reported for the search
of a dark photon with invisible decays.

\begin{figurehere}
\bcl
	\subfigure{\includegraphics[width=0.45\textwidth]{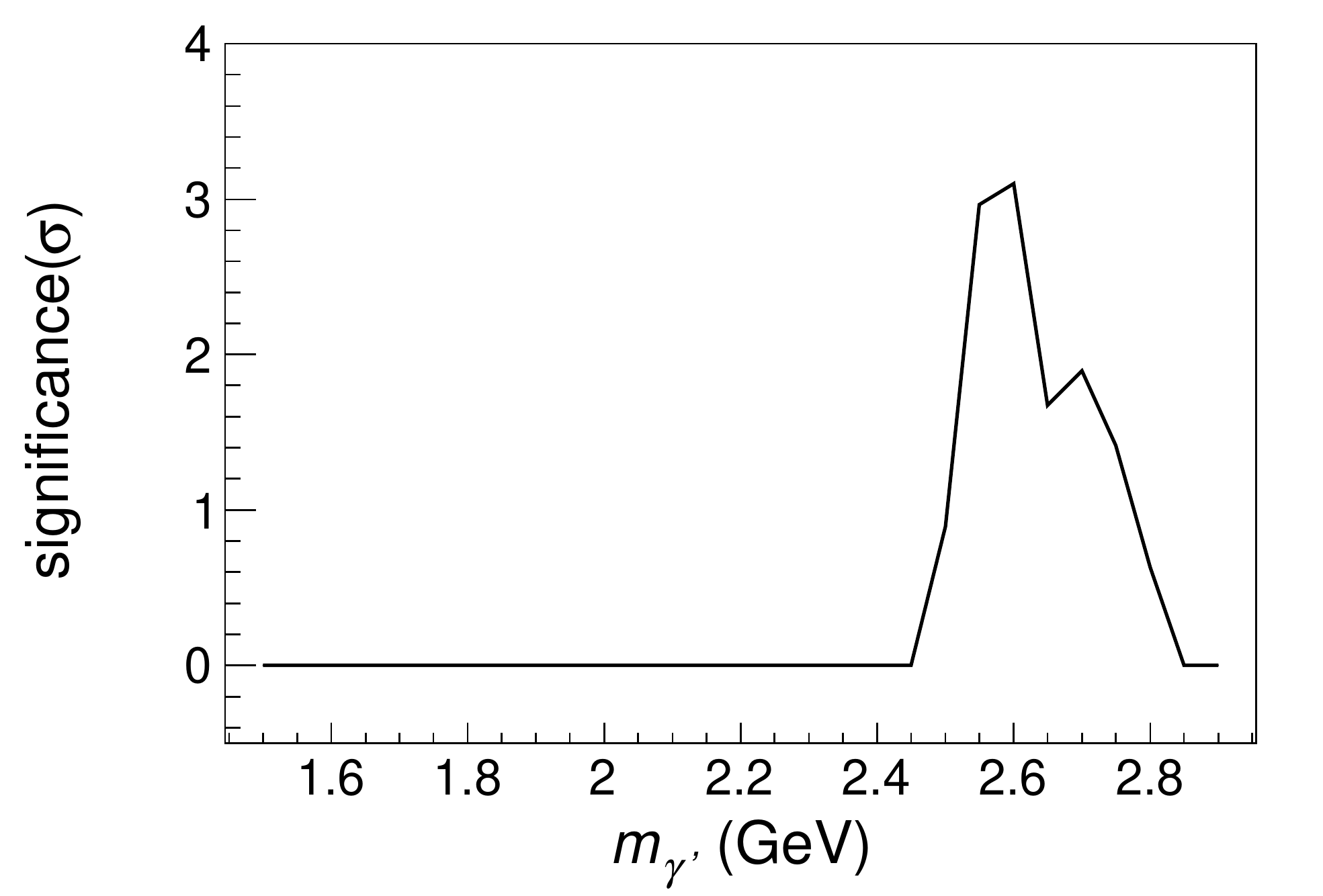}}
	\caption{Local statistical significance for dark photon
          candidates obtained from the fit to the photon energy
          spectrum as a function of
          $m_{\gamma'}$. }
	\label{fig:sig}	
\ecl
\end{figurehere}

\section{Systematic uncertainties}
The systematic uncertainties for the $\epsilon$ measurement include
those from the luminosity measurement, photon detection efficiency,
selection criteria, photon resolution, and determination of the
trigger efficiency.  The luminosity of the data sets used is measured
using large angle Bhabha scattering events, with an uncertainty less than
1.0$\%$~\cite{BESIII:2022xii}.  The signal process has only one signal
photon, and 1.0$\%$ is assigned as the uncertainty from the photon
detection efficiency~\cite{BESIII:2010ank}.

The systematic uncertainties from the requirement on the number of
photons, EMC time, $E_{\rm extra}$, EMC shower shapes, and MUC hits
are investigated with a clean di-gamma control sample. We use two
back-to-back photons with energies close to the beam energy to select
a di-gamma control sample from the same
data sets~\cite{BESIII:2022xii}.  The systematic uncertainties, which
are determined from the differences between signal MC events and
di-gamma data events, are 1.4$\%$ for the number of photons, 1.4$\%$
for the EMC time, 0.3$\%$ for $E_{\rm extra}$, 0.4$\%$ for EMC shower
shapes, and 8.5$\%$ for the MUC hits. 
The relative large systematic uncertainty from MUC hits is due to
the poor simulation of the raw informations as well as the noise from electronics.

In our fits to the photon energy spectra, the signal PDF is
convolved with a Gaussian function to account for the resolution
difference between data and MC simulation.  The parameters of the
Gaussian function are obtained from di-gamma data events with
c.m.\ energies from 2.644 to 3.400~GeV to ensure the photon energy
is similar to the signal process.
To determine the systematic uncertainty due to the photon energy
resolution, signal PDF, and background modeling, the fit is repeated with
alternative signal and background shapes, i.e.\ varying the width of
the Gaussian function used to convolve the signal shape according to its
uncertainty, replacing the signal PDF to the sum of two Crystal Ball functions 
with a common mean value, and using the $(n+1)^{th}$ order polynomial functions for
the background shape. The one with the largest upper limit value for
$\epsilon$ is taken as the final result.

For the systematic uncertainty due to the trigger efficiency, the small
trigger efficiency difference measured by electrons and photons in the
EMC barrel, which is 1.0$\%$, is taken as the systematic uncertainty.

Table~\ref{tab:sys_errors} lists all considered sources of systematic
uncertainties. Assuming they are independent, the total systematic
uncertainty is the sum in quadrature of the individual contributions and is
determined to be 8.9$\%$. Compared with the statistical uncertainty of data
which is in $\sim 30\%$ level (cf. Fig.~\ref{fig:fit_res}), the systematic uncertainty is still much smaller
and does not play a significant role in the $\epsilon$ measurement.

 \begin{tablehere}

	\caption{Sources of relative systematic uncertainties and their contributions ($\%$).}  \bcl
	\label{tab:sys_errors}
	\begin{tabular}{l    c   }
		\hline\hline
		\centering{Source} & {\centering{Uncertainty}} \\ 
		\hline
		Luminosity  &  1.0 \\
		Photon detection  & 1.0 \\
		 Number of photons    &   1.4   \\ 
		 EMC time     &   1.4   \\ 
		 $E_{\rm extra}$ & 0.3\\
		 Shower shapes & 0.4\\
		 MUC hits & 8.5\\
		 Trigger efficiency & 1.0\\ 
		 \hline	 
		 Total & 8.9\\
		\hline\hline
	\end{tabular}
\ecl	
\end{tablehere}

\section{Upper limit for coupling strength}
Since no significant dark photon signal is observed, we set an upper
limit on $\epsilon$ at the 90$\%$ confidence level (C.L.).  From
Eq.~(\ref{xs}), for a given dark photon mass, the expected signal event
yield in data depends on $\epsilon$.  Applying a Bayesian
method, a likelihood scan is performed by varying the value of $\epsilon$
in the simultaneous fit to all data sets,
\begin{linenomath}
\begin{equation}
\label{eq:uplimit}
\int_{0}^{\epsilon_{90\%}}\mathcal{L}(\epsilon) d\epsilon = 90\%\int_{0}^{\infty}\mathcal{L}(\epsilon)d\epsilon  
\end{equation}
\end{linenomath}
where $\mathcal{L}(\epsilon)$ is the $\epsilon$-dependent likelihood
value, and $\epsilon_{90\%}$ represents the coupling strength
parameter which corresponds to 90\% of the integral of the likelihood
function from $\epsilon=0$ to $\infty$. To consider the systematic
uncertainty, the likelihood curves are also convolved with a Gaussian
function with its standard deviation set to the value of the total
systematic uncertainty.  The upper limits of $\epsilon$ at the 90\%
C.L.\ versus the dark photon mass are shown in Fig.~\ref{fig:upper}, and
$\epsilon_{90\%}$ varies within (1.6 - 5.7)$\times10^{-3}$, depending
on $m_{\gamma'}$ between 1.5 and 2.9~GeV.

\begin{figurehere}
\bcl
	\subfigure{
		\includegraphics[width=0.45\textwidth]{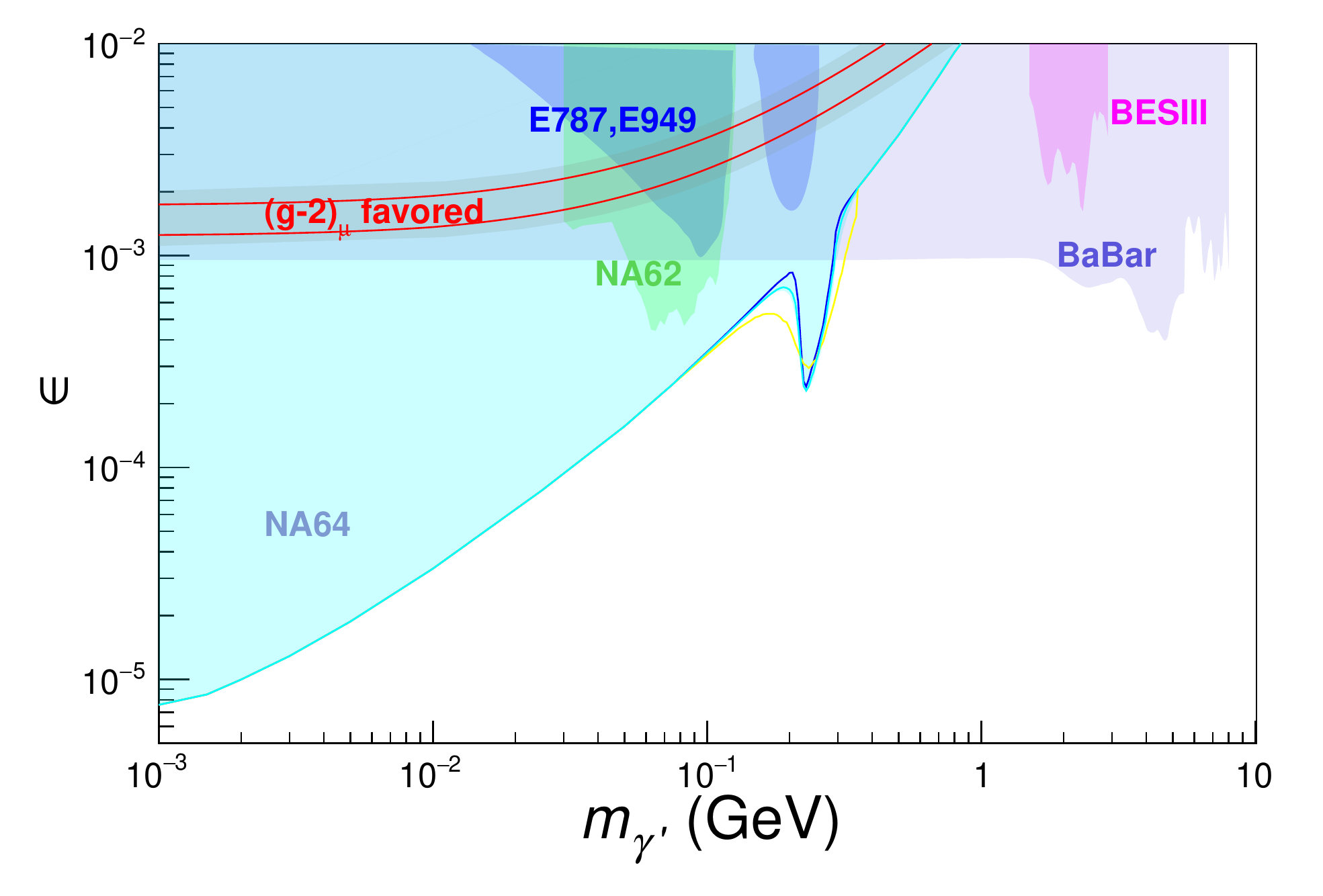}		
	}
	\caption{Upper limit on the coupling strength ($\epsilon$)
          between the dark sector and the SM at the 90\% C.L.  versus
          the dark photon mass measured by BESIII (magenta region),
          together with previous measurements~\cite{NA62:2019meo,
            NA64:2016oww, Andreev:2021fzd, BaBar:2017tiz,
            E787:2001urh, BNL-E949:2009dza}, as well as the parameter
          region favored by the $(g-2)_{\mu}$
          anomaly~\cite{Fayet:2007ua}. The two red lines correspond
          to $(g-2)_{\mu}+ 2\sigma$ and $(g-2)_{\mu}-2\sigma$ and
          take into account the latest result from
          Fermilab~\cite{Muong-2:2021ojo}.}
	\label{fig:upper}	
\ecl	
\end{figurehere}

\section{Summary} In summary, using data collected 
at c.m.\ energies from 4.13 to 4.60 GeV corresponding to an
integrated luminosity of 14.9 fb$^{-1}$, we search for the radiative annihilation
production of a dark photon which decays invisibly at the BESIII experiment 
for the first time. No obvious signal is
observed in the mass region between 1.5 and 2.9~GeV/$c^2$, and the
upper limit on the coupling strength parameter $\epsilon$ at the 90\%
C.L.\ is determined to be within (1.6 - 5.7)$\times10^{-3}$ as a
function of the dark photon mass. 
Our exclusion limits are below the
$(g - 2)_{\mu}$ anomaly values and are consistent with what already excluded 
by BaBar~\cite{BaBar:2017tiz} in the mass range between 1.5 and 2.9~GeV.
It also proves the capability of the BESIII detector to produce more competitive results with the coming 17~fb$^{-1}$ data taken at 3.77~GeV~\cite{Ablikim:2019hff}.

The BESIII collaboration thanks the staff of BEPCII and the IHEP computing center for their strong support. This work is supported in part by National Key R\&D Program of China under Contracts Nos. 2020YFA0406400, 2020YFA0406300; National Natural Science Foundation of China (NSFC) under Contracts Nos. 11635010, 11735014, 11835012, 11935015, 11935016, 11935018, 11961141012, 12022510, 12025502, 12035009, 12035013, 12192260, 12192261, 12192262, 12192263, 12192264, 12192265; the Chinese Academy of Sciences (CAS) Large-Scale Scientific Facility Program; Joint Large-Scale Scientific Facility Funds of the NSFC and CAS under Contract No. U1832207; CAS Key Research Program of Frontier Sciences under Contract No. QYZDJ-SSW-SLH040; 100 Talents Program of CAS; Project ZR2022JQ02 supported by Shandong Provincial Natural Science Foundation; INPAC and Shanghai Key Laboratory for Particle Physics and Cosmology; ERC under Contract No. 758462; European Union's Horizon 2020 research and innovation programme under Marie Sklodowska-Curie grant agreement under Contract No. 894790; German Research Foundation DFG under Contracts Nos. 443159800, Collaborative Research Center CRC 1044, GRK 2149; Istituto Nazionale di Fisica Nucleare, Italy; Ministry of Development of Turkey under Contract No. DPT2006K-120470; National Science and Technology fund; STFC (United Kingdom); The Royal Society, UK under Contracts Nos. DH140054, DH160214; The Swedish Research Council; U. S. Department of Energy under Contract No. DE-FG02-05ER41374.

\normalem

\end{multicols}
\end{document}